\documentstyle[preprint,aps]{revtex}

\begin{document}

\draft
\preprint{LBL-45309}

\title{Dynamical Screening and Radiative Parton Energy 
Loss in a Quark-gluon Plasma}

\author{Xin-Nian Wang}
\address{Nuclear Science Division, Mailstop 70-319, 
        Lawrence Berkeley Laboratory\\
University of California, Berkeley, California 94720. }
\date{\today}
\maketitle
\begin{abstract}
Dynamical screening in the magnetic part of the one-gluon exchange interaction
is included in the study of radiative energy loss of a fast parton propagating
inside a quark-gluon plasma. As a result the final radiative energy loss is
about twice as large as when only the electric part of one-gluon exchange 
interaction is
considered. A non-perturbative magnetic screening mass is also used in the 
estimate of the mean-free-path of parton scattering in a hot QCD matter.  
\end{abstract}


Radiative energy loss of a fast parton inside a hot QCD matter has been 
proposed as a good probe of the medium and should lead to observable 
consequences such as jet quenching in high-energy heavy-ion 
collisions \cite{quenching0,quenching1,quenching2,quenching3}. 
Theoretical estimate of the radiative
energy loss suffered by a fast parton in a hot QCD medium has attracted
a lot of interests because it helps us to understand the dependence of
the energy loss on the properties of the medium and in particular the
difference between parton energy loss in a cold nuclear medium and a hot
quark-gluon plasma. 

Radiative energy loss has been estimated in various approaches, from 
uncertainty principle analysis \cite{brodsky} to calculation of induced 
radiation in a multiple scattering model \cite{gw}. 
A very interesting feature of the radiative energy loss found by
a recent study in Ref.~\cite{BDMPS}, referred to as BDMPS in this
paper, is that the energy loss depends quadratically on the distance
that the parton travels through. BDMPS demonstrated that such a nonlinear
dependence arises from the non-abelian gluon rescattering in the medium.
Most of these studies used the screened static-potential model for multiple
scattering in a hot medium as proposed by Gyulassy and Wang (GW) \cite{gw}.
In the GW model for multiple scattering, the interaction suffered by the
propagating parton is assumed to be by a static potential with Debye
screening. Such a screened static potential model gives finite cross
section and average transverse momentum broadening. Even though BDMPS
and Zakharov \cite{zak,zak2} 
later on generalized the study to other models
of parton scattering, the problem of the magnetic part of one-gluon
exchange interaction in a medium and its effect on the radiative 
energy loss remains unexplored.

In this paper we will study the radiative energy loss of a fast parton
inside a quark-gluon plasma including both the electric and magnetic part
of the strong interaction. The magnetic part of the one-gluon exchange 
interaction is not screened perturbatively in the static limit in a hot QCD 
plasma. One therefore has to introduce a non-perturbative magnetic screening
mass $\mu_{\rm mag}$ 
in order to calculate the parton scattering cross section or
the mean-free-path of a propagating parton similar to the calculation of the
gluon damping rate \cite{hw96}. For the calculation of some transport
quantities, like the average momentum transfer per interaction, the dynamical
screening provided by the imaginary part of the self-energy in the
magnetic interaction is enough to regulate the infrared behavior of
the magnetic interaction and gives finite results. In both cases, 
correlation scales provided by the static and dynamics magnetic 
screening are somewhat different from the static electric screening. 
They should have significant effect on radiative parton energy loss
in a quark-gluon plasma.

According to BDMPS \cite{BDMPS}, 
the radiative energy loss of a fast parton inside a
medium with finite size $L$ is
\begin{equation}
\frac{dE}{dz}=\frac{\alpha_s N_c}{4} \langle p^2_{\perp W}\rangle,
\end{equation}
for any model of multiple parton scattering, where
$\langle p^2_{\perp W}\rangle$ is the total accumulated momentum broadening
during the parton's propagation inside the medium which grows linearly with
the media length $L$, {\it i.e.,} 
$\langle p^2_{\perp W}\rangle =L  d\langle p^2_{\perp}\rangle$/dL. The momentum
broadening per unit distance is
\begin{equation}
\frac{d\langle p^2_{\perp}\rangle}{dL}= \rho \int_0^{\mu^2/B^2} dq^2 
q^2 \frac{d\sigma}{dq^2} \, , \label{eq:pt1}
\end{equation}
where $\rho$ is the media parton density, $B=\lambda/L$ with $\lambda$ 
being the mean-free-path of the
propagating parton and $\mu$ is the typical momentum transfer in
a parton scattering which is the Debye screening mass $\mu_D$ in the
GW model of multiple scattering. In a hot quark-gluon plasma, one
should include both the electric and magnetic interaction of
one-gluon exchange. One should also replace Eq.~(\ref{eq:pt1}) with its
thermal averaged value,
\begin{equation}
\frac{d\langle p^2_{\perp}\rangle_a}{dL}=
\sum_b\nu_b\int \frac{d^3p_b}{(2\pi)^3} f(p_b)(1\pm f(p_c)) \frac{d^3p_c}{(2\pi)^3}
\frac{d^3p_d}{(2\pi)^3}\, q^2 |{\cal M}_{ab}|^2 (2\pi)^4\delta^4(p+p_b-p_c-p_d)
\label{eq:pt2}
\end{equation}
where we use the index $a$ to denote the flavor of the fast parton and $f(p)$
is the Bose-Einstein $f_{BS}$ (Fermi-Dirac $f_{FD}$) 
distribution for the thermal gluons (quarks) in the medium.
We will only consider the elastic channels that are dominant at small
angles.
The statistical factor $\nu_2$ is $2(N_c^2-1)$ for
gluons and $4N_cn_f$ for $n_f$ flavors of quarks. In this paper we will
assume $n_f=2$. We neglect the quantum statistical effect for the
fast partons. We denote the energy and momentum transfer of the parton
scattering by $\omega$ and $q$, respectively. The above integral is 
dominated by contributions from small angle scattering. 
In this small-angle approximation, i.e., $\omega, q \ll E, E_b$,
energy-momentum conservation leads to
\begin{eqnarray}
\vec{p}_c=\vec{p}+\vec{q}\, , \,\,\,\, \vec{p}_d=\vec{p_b}-\vec{q} \nonumber \\
E_c=E+\omega \, , \,\,\,\, E_d=E_b-\omega \nonumber \\
\omega \approx \vec{v}\cdot\vec{q}\approx \vec{v}_b\cdot\vec{q},
\end{eqnarray}
where $\vec{v}=\vec{p}/E$ and $\vec{v}_b=\vec{p}_b/E_b$.
In the small-angle scattering limit, the effective
matrix element for parton scattering is \cite{hw96},
\begin{equation}
{\cal M}_{ab}\approx g^4C_{ab}\left[ \frac{1}{q^2+\mu_D^2\pi_L(x)}
-\frac{(1-x^2)\cos\phi}{q^2(1-x^2)+\mu_D^2\pi_T(x)}\right]\, ,
\end{equation}
where $\cos\phi=(\vec{v}\times\vec{q})\cdot(\vec{v}_b\times\vec{q})/q^2$
, $x=\omega/q$ and $\mu_D^2=g^2(N_c+n_f/2)T^2/3$ is the Debye screening 
mass in thermal QCD medium with temperature $T$. 
The color factors for parton scattering are
$C_{qq}=(N_c^2-1)/4N_c^2=2/9$, $C_{qg}=1/2$ and $C_{gg}=N_c^2/(1-N_c^2)=9/8$.
We use an effective gluon propagator
to include the resummation of an infinite number of loop corrections 
\cite{pisarski}. The scaled self-energies in the effective propagator
in the long-wavelength limit are given by \cite{weldon},
\begin{eqnarray}
  \pi_L(x)&=&1-\frac{x}{2}
  \ln\left(\frac{1+x}{1-x}\right) + i \frac{\pi}{2}x 
  \, , \label{eq:self1}\\
  \pi_T(x)&=&\frac{x^2}{2}+\frac{x}{4}(1-x^2)
  \ln\left(\frac{1+x}{1-x}\right) - i \frac{\pi}{4}x(1-x^2)
  \, .\label{eq:self2}
\end{eqnarray}

In Eq.~(\ref{eq:pt2}), the integration over 
$p_c$ and $p_d$ can be rewritten as
\begin{equation}
(2\pi)^4\int \frac{d^3p_c}{(2\pi)^3}\,\frac{d^3p_d}{(2\pi)^3}
\delta^4(p+p_b-p_c-p_d)
=\frac{1}{(2\pi)^2}\int d^3q\int_{-q}^{q} d\omega 
\delta(\omega-\vec{v}\!\cdot\!\vec{q})
\delta(\omega-\vec{v}_b\!\cdot\!\vec{q})\; .
\end{equation}
The two $\delta$-functions will fix the angular integrals of $p_b$ and $q$.
With approximation $f(p_c)\approx f(p_b)$ and the integrals
\begin{eqnarray}
\int dp\,p^2f_{BS}(p)(1+f_{BS}(p))&=&T^3\pi^2/3\, , \nonumber \\ 
\int dp\,p^2f_{FD}(p)(1-f_{FD}(p))&=&T^3\pi^2/6\, ,
\end{eqnarray}
we obtain the
averaged momentum transfer per unit distance in Eq.~(\ref{eq:pt2}) as
\begin{eqnarray}
 \frac{d\langle p^2_{\perp}\rangle_a}{dL} &=&\frac{g^4}{2\pi}C_a T^3
  \int_{-1}^1dx\int_0^{q^2_{\rm max}/B^2}\frac{dq^2q^2}{\mu_D^4} \nonumber \\
&\times&\left\{\frac{1}{|q^2/\mu_D^2+\pi_L(x)|^2}+
\frac{1}{2}\frac{(1-x^2)^2}{|(1-x^2)q^2/\mu_D^2+\pi_T(x)|^2}\right\}\; ,
\label{eq:pt3}
\end{eqnarray}
where $C_q=4/9$ and $C_g=1$. The integration over $p_b$ provides a cut-off
for $q$ integration at $q_{\rm max}^2=3ET/2$. 
The integral in Eq.~(\ref{eq:pt3})
should be a function of $q^2_{\rm max}/\mu_D^2B^2$ only.

From Eq.~(\ref{eq:self1}) and (\ref{eq:self2}), we can see that with 
Debye screening the longitudinal contribution to the above integral
is finite. However, the real part of the transverse self-energy
vanishes quadratically in the static limit ($x\rightarrow 0$). Without the
imaginary part this would have caused a quadratical divergency in the
transverse propagator.  Fortunately, imaginary part provides Landau damping to
parton interactions in a thermal medium and reduce the divergency to
a logarithmic singularity in the static limit. When weighted with the
momentum transfer $q^2$, the transverse contribution to the 
integral is then finite. A fit to the numerical evaluation of the integral
gives,
\begin{eqnarray}
{\cal I}_L&=&0.92 \ln\frac{q^2_{\rm max}}{\mu_D^2B^2} , \nonumber \\
{\cal I}_T&=&0.5 \ln \frac{q^2_{\rm max}}{\mu_D^2B^2}.
\end{eqnarray}
We can see that the contribution from the magnetic interaction is as big
as the electric one. The final momentum broadening per unit distance and the
radiative parton energy loss are then
\begin{eqnarray}
\frac{d\langle p^2_{\perp}\rangle_a}{dL}&=&
8\pi C_a T^3 \alpha_s^2 1.42\ln\frac{3ETL^2}{2\mu^2_D\lambda_a^2}\, ,
\nonumber \\
\frac{dE_a}{dz}&=& \frac{N_c\alpha_s}{4}
L\frac{d\langle p^2_{\perp}\rangle_a}{dL} . \label{eq:de1}
\end{eqnarray}

In order to complete the estimate of the radiative energy loss, we now
have to estimate the mean-free-path of parton scattering in the same
framework. Similarly including both the electric and magnetic part of
interaction, one has
\begin{eqnarray}
 \lambda_a^{-1}\equiv\langle \rho\sigma\rangle_a &=&\frac{g^4}{2\pi}
C_a \frac{T^3}{\mu_D^2}
  \int_{-1}^1dx\int_0^{q^2_{\rm max}}\frac{dq^2}{\mu_D^2} \nonumber \\
&\times&\left\{\frac{1}{|q^2/\mu_D^2+\pi_L(x)|^2}+
\frac{1}{2}\frac{(1-x^2)^2}{|(1-x^2)q^2/\mu_D^2+\mu_{\rm mag}^2
+\pi_T(x)|^2}\right\}\; .
\label{eq:lambda}
\end{eqnarray}
Unlike in Eq.~(\ref{eq:pt3}) for the averaged momentum transfer, the 
logarithmic singularity in the magnetic part of the gluon propagator
can only be regularized by introducing a non-perturbative magnetic screening
mass $\mu_{\rm mag}\sim g^2T$ much like in the calculation of the 
damping rate of a
fast parton \cite{pisarski,hw96}. In the weak coupling limit, the dominant
contribution in the magnetic interaction comes from 
$\mu_{\rm mag}\stackrel{<}{\sim} q \stackrel{<}{\sim} \mu_D$ and is 
independent of the cut-off $q_{\rm max}\gg \mu_D$. Numerically one 
can carry out the above integration and find the integral as
\begin{eqnarray}
{\cal I}_{\lambda}&=&2(\ln\frac{\mu_D^2}{m_{\rm mag}^2}-1.0
  +2.0\frac{\mu_{\rm mag}^2}{q_D^2} - 0.32\frac{\mu_D^2}{q_{\rm max}^2})
  + 2.2\frac{q_{\rm max}^2}{q_{\rm max}^2+\mu_D^2} \nonumber\\
&\approx&2(\ln\frac{\mu_D^2}{m_{\rm mag}^2}-0.1) +{\cal O}(g^2) \; ,
\end{eqnarray}
where the first term is the magnetic and the second term is the 
electric contribution. Using the estimate of 
$\mu_{\rm mag}\approx 0.255\sqrt{N_c/2}g^2T$ from Ref.~\cite{TBBM93},
we have
\begin{equation}
\lambda_a^{-1}=3TC_a\alpha_s\ln\frac{1}{\alpha_s},
\end{equation}
with the dominant contribution from the magnetic interaction. In the weak 
coupling limit, the magnetic contribution significantly changes the 
mean-free-path of a propagating parton. However, for a practical value
of $\alpha_s=1/3$, the net result is about the same as when only the
electric interaction with Debye screening is considered.
Substitute the above mean-free-path into Eq.~(\ref{eq:de1}), we have 
the parton energy loss in a quark-gluon plasma,
\begin{equation}
\frac{dE_a}{dz}=4\pi N_c\alpha_s^3T^3L C_a1.42
\ln\left[\frac{9}{2}C_aL\ln\frac{1}{\alpha_s}\sqrt{2\pi\alpha_sET}\right].
\end{equation}

For a fast quark with $E=250$ GeV traveling through a hot quark-gluon plasma 
with $T=250$ MeV and $L\simeq 10 $ fm we have
\begin{eqnarray}
\frac{d\langle p_T^2\rangle_a}{dL}&\approx& 1.9 
\;\;{\rm GeV}^2/{\rm fm} \nonumber \\
\frac{dE}{dz}& \approx& 24 \;\;{\rm GeV/fm} \left(\frac{L}{10 \,{\rm fm}}\right)
\, .
\end{eqnarray}
This is about 4 times larger than the original estimate by
BDMPS \cite{BDMPS,zak2}. 
It is in part due to the contribution from the magnetic
interaction that was not considered before and in part due to the different 
estimate of the mean-free-path in this paper which depends on the 
non-perturbative magnetic screening mass $\mu_{\rm mag}$. For a smaller
value of $\alpha_s=1/10$ that is more relevant for the week coupling limit in
our estimate, we have
\begin{eqnarray}
\frac{d\langle p_T^2\rangle_a}{dL}&\approx& 0.18
\;\;{\rm GeV}^2/{\rm fm} \nonumber \\
\frac{dE}{dz}& \approx& 0.68 \;\;{\rm GeV/fm} \left(\frac{L}{10 \,{\rm fm}}\right)
\end{eqnarray}

To conclude, we have considered both the electric and magnetic part of
one-gluon exchange interaction in the estimate of radiative energy loss
by a fast parton propagating in a hot QCD plasma. We used the results by 
BDMPS \cite{BDMPS} for the energy loss which is proportional to the
averaged momentum transfer per unit distance. The imaginary part of
the magnetic self-energy which is responsible for Landau damping regularizes
the collisional integral in the calculation and gives a finite averaged 
momentum transfer. We found the contribution from the magnetic interaction
is as big as the electric interaction. In the estimate of the mean-free-path,
we have to introduce a magnetic screening mass which gives an additional
logarithmic dependence on the strong coupling constant $\ln 1/\alpha_s$. The
final radiative energy loss has a cubic dependence on the coupling constant.
Because of our treatment of the magnetic interaction, our final numerical 
estimate of radiative energy loss is about 4 times larger as the original
estimate by BDMPS.

\acknowledgments
This work was supported by the Director, Office of Energy 
Research, Office of High Energy and Nuclear Physics,
Division of Nuclear Physics, and by the Office of Basic Energy Science, Division
of Nuclear Science, of the U.S. Department of Energy 
under Contract No. DE-AC03-76SF00098. It is also supported in part by
NSFC under project 19928511.

\end{document}